# Deployment of VoIP Technology: QoS Concerns

Amor Lazzez [1], Thabet Slimani [2]

Assistant Professor, Taif University, Kingdom of Saudi Arabia[1, 2]

**Abstract**: Voice over IP (VoIP) is an emerging communication service allowing voice transmission over a private or a public IP network. VoIP allows significant benefits for customers and service providers including cost savings, phone and service portability, mobility, and the integration with other applications. Nevertheless, the deployment of the VoIP technology encounters many challenges such as interoperability issues, security issues, and QoS concerns. Among these disadvantages, QoS issues are considered the most serious due to the QoS problems that may arise on IP networks, and the stringent QoS requirements of voice traffic. The aim of this paper is carry out a deep analysis of the QoS concerns of the VoIP technology. Firstly, we present a brief overview about the VoIP technology. Then, we discuss the QoS issues related to the use of the IP networking technology for voice traffic transmission. After that, we present the QoS concerns related voice clarity. Finally, we present the QoS mechanisms proposed to make the IP technology able to support voice traffic QoS requirements in terms of voice clarity, voice packet delay, packet delay variation, and packet loss.

## I. INTRODUCTION

Voice over IP (VoIP) [1, 2, 10, 11] has been prevailing in the telecommunication world since its emergence in the late 90s, as a new technology transporting multimedia over the IP network. The reason for its prevalence is that, compared to legacy phone system, VoIP allows significant benefits such as cost savings, the provision of new media services, phone portability, and the integration with other applications [1, 2, 10, 11]. Despite these advantages, the VoIP technology suffers from many hurdles such as architecture complexity, interoperability issues, QoS concerns, and security issues [1, 2, 7, 8, 10, 11]. Among these disadvantages, QoS issues are considered the most serious for the two reasons [7-10]. The first is that voice traffic is characterized by stringent QoS requirements in terms of voice clarity, packet delay, packet delay variation, and packet loss [7, 10, 13, 15]. The second reason is that the IP networking technology is limited to the provision of a Best Effort service which makes it enable to support with strict QoS requirements like voice and video traffic [7, 8, 10].

Because of the nature of the IP networking technology, data packets sent via an IP network are subject to certain transmission problems such as packet delay, packet delay variation (jitter), packet loss [7-11]. To overcome these transmission problems, and hence make the IP technology able to support emerging multimedia applications with stringent QoS constraints, QoS mechanisms have been considered [1,2, 7-11].

QoS is defined as the ability of the network to provide better or "special" service to selected users and applications, to the potential detriment of other users and applications [7, 14]. The deployment of the QoS aspect helps the provision of bandwidth guarantees while minimizing network delay, jitter, and loss ratio for prioritized traffic like voice traffic. They do so not by creating additional bandwidth, but by controlling how the available bandwidth is used by the different applications and protocols on the network. In effect, this often means that data applications and protocols are restricted from accessing bandwidth when VoIP traffic needs it. This does not have much of an impact on the data traffic, however, because it is generally not as delay or drop-sensitive as VoIP.

Voice traffic is very sensitive to delayed packets, lost packets, and variable delay [7, 10, 13, 15, 16]. The effects of these problems manifest as choppy audio, missing sounds, echo, or unacceptably long pauses in the conversation that cause overlap, or one talker interrupting the other [7, 14]. To help an efficient deployment of the VoIP technology, QoS mechanisms making the IP technology able to support the stringent QoS constraints of voice traffic have been considered. Actually QoS mechanisms allowing the increase of the available bandwidth on an IP network, as well as the decrease of packet delay, jitter, and packet loss have been developed [7-8, 12-16].

In addition to the QoS concerns that are related to the use of the IP networking technology to transmit voice traffic, a number of concerns related to voice clarity should be considered to help the deployment of a successful VoIP system [7,8, 12,13]. The clarity of the audio signal is of highest importance. The listener must be able to recognize the identity and sense the mood of the speaker. Voice clarity can be affected by different factors including fidelity, echo, and side tone, and the background noise [7,8].

The remaining of this paper is organized as follows. Section 2 presents an overview about VoIP architectures and protocols. First, we present the VoIP architecture. Then, we highlight the benefits leading to the ever-growing of the

VoIP popularity. After that, we present a brief overview about the main VoIP protocols. Section 3 presents the QoS requirements of voice traffic, highlights the QoS problems that may arise on IP networks, and present the considered QoS mechanisms to make the IP technology able to support voice traffic QoS needs. Section 4 presents the QoS concerns related to voice clarity including fidelity, echo, side tone, and background noise. Section 5 concludes the paper.

## II. BASICS OF VoIP TECHNOLOGY

VoIP is a rapidly growing technology that delivers voice communications over Internet or a private IP network instead of the traditional telephone lines [1, 2, 11, 16]. VoIP involves sending voice information in the form of discrete IP packets sent over Internet rather than an analog signal sent throughout the traditional telephone network. VoIP helps the provision of significant benefits for users, companies, and service providers. Cost savings, the provision of new communication services, phone and service portability, mobility, and the integration with other applications are examples of the VoIP benefits. Yet, the deployment of the VoIP technology encounters many difficulties such as architecture complexity, interoperability issues, QoS issues, and security concerns [1-4].

In the following subsections, we first we highlight the benefits of the VoIP technology leading to the ever-increasing of its popularity. Then, we present the main architecture used in the deployment of the VoIP technology. After that, we present a brief overview the most important VoIP protocols. Finally, we mention the main concerns of the VoIP technology.

### A. VoIP Benefits

The key benefits of the VoIP technology are as follows [1, 2, 10, 11]:

*Cost savings:* The most attractive feature of VoIP is its cost-saving potential. Actually, for users, VoIP makes long-distance phone calls inexpensive. For companies, VoIP reduces cost for equipment, lines, manpower, and maintenance. For service providers, VoIP allows the use of the same communication infrastructure for the provision of different services which reduces the cost of services deployment.

*Provision of new communication services:* In addition to the basic communications services (phone, fax), the VoIP technology allows users to check out friends' presence (such as online, offline, busy), send instant messages, make voice or video calls, and transfer images, and so on.

*Phone portability:* VoIP provides number mobility; the phone device can use the same number virtually everywhere as long as it has proper IP connectivity. Many businesspeople today bring their IP phones or soft-phones when traveling, and use the same numbers everywhere.

*Service mobility:* Wherever the user (phone) goes, the same services will be available, such as call features, voicemail access, call logs, security features, service policy, and so on.

*Integration and collaboration with other applications:* VoIP allows the integration and collaboration with other applications such as email, web browser, instant messenger, social-networking applications, and so on.

### B. Client-Server VoIP Architecture

One of the main features of the VoIP technology is that it may be deployed using a centralized or a distributed architecture [1, 2, 10, 11, 16]. The majority of current VoIP systems are deployed using a client-server centralized architecture. A client-server VoIP system relies on the use of a set of interconnected central servers known as gatekeepers, proxy servers, or soft-switches. The central servers are responsible for users' registration as well as the establishment of VoIP sessions between registered users. Figure 1 shows an example of a VoIP system deployed using the client-server architecture. As it is illustrated in the figure, each central server handles (registers, establishes a session with a local or a distant user, etc.) a set of users. Each user must be registered on one of the central servers (registrar server) to be able to exchange data with other registered users. A user gets access to the service only over the registrar server.

### C. VoIP Protocols

The deployment of any multimedia application such as VoIP, videoconference, or network gaming requires a signaling protocol to set up sessions between end points, and a distinct protocol to transmit the media streams. The standard protocol used to exchange media streams between the endpoints of an established session.

#### 1. VoIP Media Transport Protocols

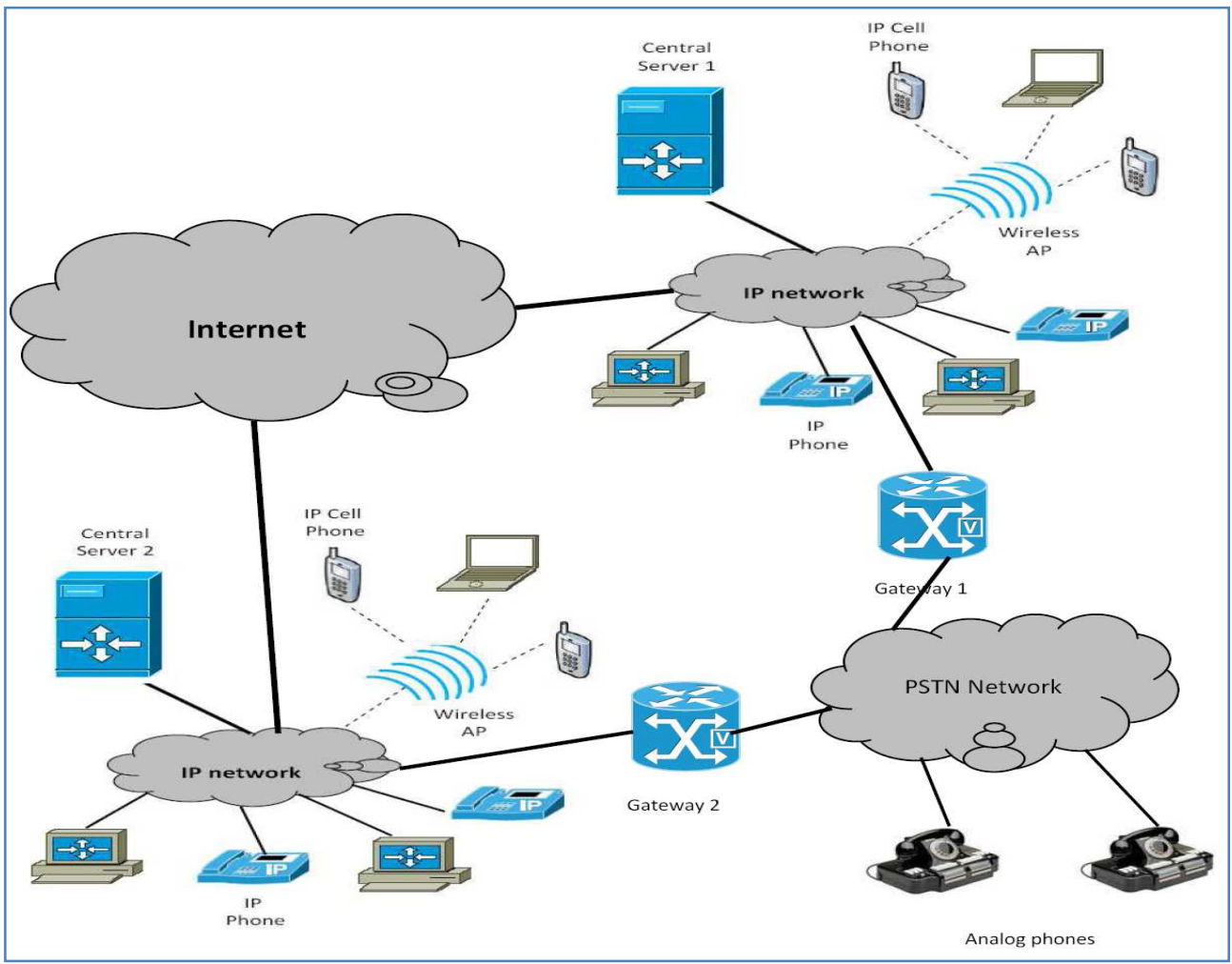

Figure1: Client-Server VoIP Architecture: An illustrative example

The majority of VoIP systems rely on the use of the Real-Time Transport Protocol (RTP) for data transmission during a VoIP session. Secure RTP (SRTP) has been recently proposed by the IETF as a secured version of the RTP protocol.

*RTP Protocol:* Defined in RFC 3550, RTP protocol defines a standardized packet format for delivering audio and video over IP networks [1-2]. RTP is used in conjunction with

*RTP Control Protocol (RTCP):* While RTP carries the media streams (audio and video), RTCP monitors transmission statistics and the provided QoS and aids synchronization of multiple streams.

*SRTP Protocol:* SRTP protocol defines a security profile of RTP, intended to provide the authentication, the confidentiality, and the integrity of RTP messages [1, 2]. Since RTP is used in conjunction with RTCP, SRTP is closely related to SRTCP (Secure RTCP) which is used to control the SRTP session.

## 2. VoIP Signaling Protocols

The main signaling protocols considered in deployment of client-server VoIP systems are H323, SIP, and IAX.

*H323:* Standardized by the International Telecommunication Union (ITU), H323 [2] is the first signaling approach publicly used for the deployment of VoIP systems in conjunction with RTP protocol. H323 standard encompasses many protocols such as H225, H245, and H235. H.225 defines call setup messages and procedures used to establish a call, as well as messages and procedures used for users registration, and call admission. H.245 defines control messages and procedures used to exchange communication capabilities such as the supported codec. H235defines security profiles for H.323, such as authentication, message integrity, signature security, and voice encryption.

*SIP:* Allowing system flexibility and security, SIP is nowadays the most used VoIP signaling protocol [3]. SIP is an application layer protocol that works in conjunction with several other application layer protocols that identify and carry the session media. Media identification and negotiation is achieved with the Session Description Protocol (SDP). Media streams (voice, video) are transmitted using RTP protocol which may be secured by the SRTP protocol. For secure transmissions of SIP messages the Transport Layer Security (TLS) may be used. SIP also provides a suite of security services including DoS prevention, authentication, integrity, and confidentiality.

*IAX:* Currently, IAX (Inter-Asterisk Exchange) is one of the most used approaches for the deployment of VoIP systems [4-6]. In contrast with H323 and SIP protocols which are limited to signaling tasks, IAX protocol ensures both signaling and media transmission in an IAX-based VoIP system. IAX provides a suite of security services. Actually, it allows message authentication and confidentiality, and supports NAT traversal.

*D. VoIP Disadvantages*

Even though it allows significant benefits, the VoIP technology suffers from many hurdles [1, 2, 7, 8, 10, 11]. In the following a brief presentation of the main VoIP disadvantages.

*Complicated service and network architecture:* the integration of different services (voice, video, data, and so on) into the same network makes it difficult the design of the network architecture because different protocols and devices are involved for each service, and various characteristics are considered for each media. It also causes various errors and makes it harder to troubleshoot and isolate them.

*Interoperability issues:* Different protocols (H323, SIP, IAX, and MGCP) have been proposed for the deployment of VoIP systems. This leads to an interoperability issues between the VoIP devices developed based on different protocols. Interoperability issues still come up between products using the same protocol due to the multitude of protocol versions, and the ways of implementation.

*Security issues:* In the legacy phone system, the main security issue is the interception of conversations that require physical access to phone lines. In VoIP security issues are much more than that. Actually, in VoIP systems many elements (IP phones, access devices, media gateways, proxy servers, and protocols) are involved in setting up a call and transferring media between two endpoints. Each element has vulnerable factors that may be exploited by attackers to carry out security attacks.

*Quality of service (QoS) issues:* The QoS aspect was not much considered when the IP technology was designed. That is why; the IP technology remains inefficient to support traffic with stringent QoS constraints like voice and video traffic.

Among the above presented disadvantages, QoS concerns are considered the most serious. In the following, we first present the QoS problems that are related to the use of the IP networking technology for voice traffic transmission. Then, we present QoS concerns related to voice clarity including fidelity, echo, side tone, and background noise.

III. VoIP QoS Concerns

Because of the nature of the IP technology, data packets sent via an IP network are subject to certain transmission problems such as packet delay, jitter, and packet loss [7-11]. To overcome these transmission problems, and thus make the IP technology able to support emerging multimedia applications with stringent QoS constraints, QoS mechanisms should be deployed.

Voice traffic is very sensitive to delayed packets, lost packets, and variable delay. The effects of these problems manifest as choppy audio, missing sounds, echo, or unacceptably long pauses in the conversation that cause overlap, or one talker interrupting the other [7, 8, 14]. For that reason, QoS is considered as the most important feature to deploy a successful VoIP system.

QoS is defined as the ability of the network to provide better or "special" service to selected users and applications, to the potential detriment of other users and applications [7, 14]. The deployment of the QoS aspect helps the provision of bandwidth guarantees while minimizing delay and jitter for priority traffic like voice traffic. They do so not by creating additional bandwidth, but by controlling how the available bandwidth is used by the different applications and protocols on the network. In effect, this often means that data applications and protocols are restricted from accessing bandwidth when VoIP traffic needs it. This does not have much of an impact on the data traffic, however, because it is generally not as delay or drop-sensitive as VoIP traffic.

The main issues that should be addressed by the QoS aspect to adequately transport voice traffic over an IP network are the following: bandwidth, network delay, delay variation, and traffic loss [10, 11, 14]. In the following of this section, we describe the main VoIP QoS Issues, and we show how they can be addressed to guarantee the required QoS for voice traffic.

*A. Bandwidth*

The bandwidth of a transmission media (optical fiber, coaxial cable, etc.) defines its data transmission capacity in bits/second. The bandwidth of a network path composed of different LAN and WAN links corresponds to the bandwidth of the slowest link on the path. The network link with the lowest bandwidth on a network path is often referred to as a bottleneck. Bottlenecks on a network cause congestion which results into QoS problems for voice traffic. Figure 2 presents an example of a network path between two VoIP terminals over an IP network. The figure shows the location where a bottleneck is most likely to occur on the considered path. The figure also illustrates the calculation of the path bandwidth which corresponds to the bandwidth of the IP WAN link; the slowest link on the considered path.

To adequately transport voice traffic over an IP network, and hence help the deployment of a successful VoIP system, congestion should be avoided. This can be achieved using several ways including the increase of the bandwidth, traffic prioritization, and traffic compression [7-15].

**1. Link capacity increasing**

The best way to increase bandwidth is to increase the link capacity to accommodate all applications and users, with some extra bandwidth. Even though this solution sounds simple, increasing link capacity is expensive and needs time to be implemented. Fortunately, various QoS mechanisms can be used to effectively increase available bandwidth for priority applications.

**2. Delay-sensitive traffic prioritization**

It consists to:
  **a.** Classify traffic into different classes according to their QoS constraints in terms of delay, jitter, and packet loss.
  **b.** Assign a priority level to each traffic class. The highest priority level is assigned to the traffic class or real-time applications including VoIP, and videoconference.
  **c.** Ensure priority-based traffic forwarding through the network. Assigned a high priority level, real-time applications such as VoIP can get sufficient bandwidth to support their QoS requirements; voice traffic will get prioritized forwarding; and the least-important traffic will get whatever unallocated bandwidth is remaining.

**3. Traffic compression**

Different techniques have been proposed for the compression of IP traffic so that it consumes less bandwidth. We mainly distinguish payload compression, and header compression.
  - **Payload compression:** By compressing the payload of a packet, the total size is reduced. This compression method does not affect the headers, which makes it appropriate for links that require the header to be readable to route the packet correctly.
  - **Header compression:** On point-to-point IP links where the header information is not needed to route the packet, header compression may be used.

Even though, it may increase the available bandwidth, Compression takes time and CPU resources, which may increase the end-to-end network delay.

*B. Network Delay*

Network delay is the amount of time it takes a packet to travel from a source to a destination through the network. Network delay, mainly includes the processing delay, the queuing delay, the serialization delay, and the propagation delay [7, 8].
  - **Processing delay:** The time it takes a router to take a packet from an input interface and put it into the output queue of the appropriate output interface. The processing delay mainly depends on the router architecture, and the router processing speed.
  - **Queuing delay:** The time a packet resides in the output queue of a router. Due to bottlenecks, the queuing delay depends on the traffic load, the processing speed, the bandwidth of the output interface, and the queuing mechanism.
  - **Serialization delay:** The time it takes to place a packet on the physical medium for transport.
  - **Propagation delay:** The time it takes a signal to transit a media. It depends on the type of media, and the type of signal transporting the data.

Due to bottleneck conditions, improper queuing, or configuration errors, network delay may increase and hence leads to QoS issues especially for delay-sensitive applications such as VoIP. The ITU-T G.114 specification recommends that the end-to-end network delay should not exceed 150 ms [7].

Different strategies have been considered to minimize the network delay through an IP network to make the IP technology able to support real-times applications with stringent constraints in terms of delay. Network delay may be minimized using the same strategies used for the increasing the available bandwidth [7-16].

**1. Increase the transmission speed of the network links**

It helps the reduction of the serialization and transmission delays, and hence the decrease of the overall network delay.

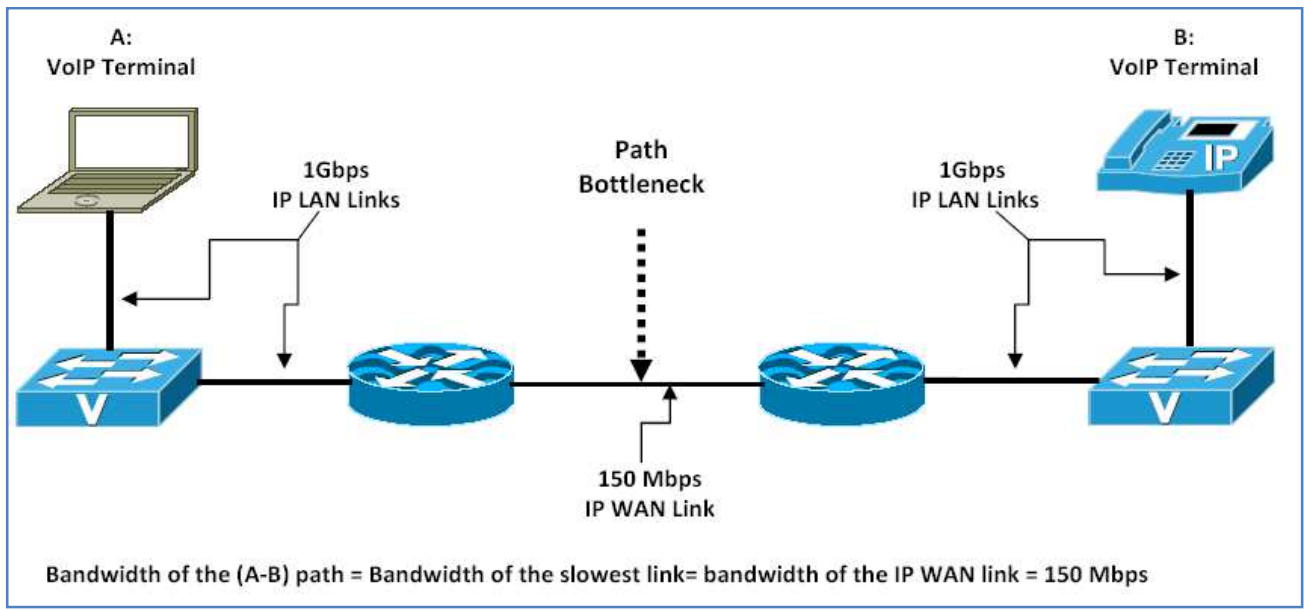

Figure 2: Network Bottleneck

**2. Increase the processing speed of the network nodes**

It allows the decrease of the processing delay, which helps the reduction of the queuing delay, and thus the decrease of the end-to-end network delay.

**3. Prioritize delay-sensitive traffic**

This approach helps the reduction of the queuing delay for delay sensitive-traffic such as voice traffic, which helps the support of the stringent delay constraints of such applications.

**4. Compress the transmitted traffic**

As it is mentioned above, traffic compressions allows the reduction of the amount of data transmitted over the network, which reduces the network traffic load, and hence decreases the queuing and the serialization delays.

*C. Delay variation*

Jitter is defined as a variation in the arrival of received packets. On the sending side, packets are sent in a continuous stream with the packets spaced evenly. Due to bottleneck conditions, this steady stream can become uneven because the delay between each packet varies instead of remaining constant. Figure 2 illustrates the jitter QoS issue.

This variation in the receiving of packets can cause the voice stream to skip and stutter, which can be very annoying to the listener. To adequately transport voice traffic over an IP network, the ITU-T G.114 specification recommends that the jitter should be reduced to 30 ms or less on average [7].

**1. De-Jitter Buffering Mechanism**

Given the annoying effects of Jitter, a QoS mechanism referred to as de-jitter or play out delay buffering has been considered [7, 8]. Implemented at the input interface of the receiving end, the de-jitter buffering mechanism relies on the use of a specific buffer known as de-jitter buffer to slow down and properly space down the received packets before being played out in a steady stream like to the transmitted one. Even though, it helps the avoidance of the jitter effects, the de-jitter mechanism affects the overall network delay.

*D. Traffic loss*

The main reason for packet loss over an IP network is network congestion. Lost data packets may be recovered by retransmission. However, lost voice packets cannot be recovered by retransmission because voice traffic must be played out in real time. Therefore QoS mechanisms minimizing voice traffic loss should be considered. For an efficient deployment of the VoIP application, The ITU-T G.114 specification recommends that the overall total of packets lost for a voice call never exceed 1 percent [7].

Voice traffic loss may be minimized using the following strategies [7-15]:
- Network congestion prevention,
- Voice traffic prioritization,
- Packet loss concealment.

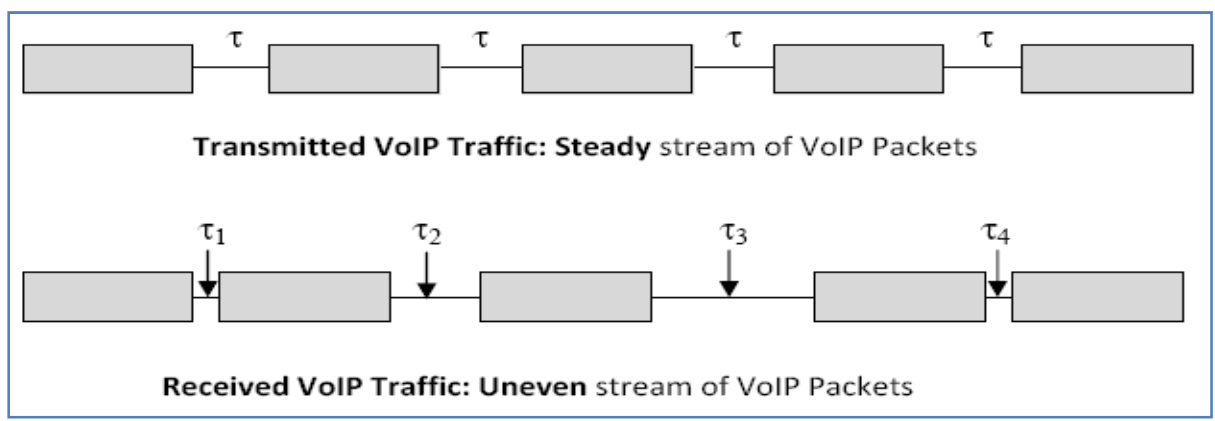

Figure 3: Jitter Issue

In the following, we present a brief overview about these strategies.

**1. Network congestion prevention**

The following procedures can be used to prevent network congestion:

- **Increase of the network bandwidth:** this can be achieved through:
  - Increase of the transmission capacity of the network links,
  - Increase of the buffering capacity of the network routers
  - Increase of the processing speed of the network routers.
- **Decrease of the traffic load:** This can be achieved through:
  - Traffic compression: it makes the data that needs to be transmitted smaller which reduces the traffic load.
  - Delay-insensitive traffic shaping: It consists to delay delay-insensitive traffic and send it at a configured maximum rate. For example, if an FTP server is generating a 512 kbps stream, shaping could limit the generated traffic to 256 kbps, delaying the transmission of the excess traffic.
  - Call Admission Control: It consists to accept a new traffic on the network only if the needed transmission bandwidth is available [12, 13].
- **Traffic policing:** it consists to drop lower-priority packets in excess a configured threshold to prevent congestion. WRED (Weighted Random Early Detection) scheme can be used start dropping these lower-priority packets before congestion occurs.

**2. Voice traffic prioritization**

It consists to delay or drop low-priority data packets to guarantee the required bandwidth for the transmission of voice traffic.

**3. Packet Loss Concealment**

Despite the above presented mechanisms which aim to reduce voice traffic loss, we cannot avoid the loss of a voice packet. The loss of a voice packet causes voice clipping and skips. As a result, the listener hears gaps in the conversation. To assist with packet loss on voice calls, a specific mechanism, referred to as Packet Loss Concealment (PLC), has been proposed. The PLC mechanism intelligently analyzes missing packets and generates a reasonable replacement packet to improve the voice quality. Cisco VoIP technology uses 20-ms samples of voice payload per VoIP packet by default. Effective codec correction algorithms require that only a single packet can be lost at any given time. If more packets are lost, the listener experiences gaps.

## IV. VOICE CLARITY CONSIDERATIONS

In addition to the QoS concerns that are related to the use of the IP networking technology to transmit voice traffic, a number of concerns related to voice clarity should be considered to help the deployment of a successful VoIP system [7,8, 12,13].

The clarity (that is, the "cleanliness" and "crispness") of the audio signal is of highest importance. The listener must be able to recognize the identity and sense the mood of the speaker. Voice clarity can be affected by different factors including fidelity, echo, and side tone, and the background noise.

*A. Fidelity*

It is defined as the degree to which the voice transmission network accurately reproduces the transmitted voice signal. Fidelity depends on the sampling frequency band and the compression ratios. When audio is sampled using the frequency band [300-3400 Hz] (narrowband) and is then highly compressed, the audio is considered to be low fidelity. However, when it is sampled using the frequency band [50–7000 Hz] (wideband) and transported using lower compression ratio is called high fidelity.

The human voice covers the wide frequency band (50-7000 Hz). Voice sampling using the narrowband allows an acceptable QoS given that 90 percent of the most significant elements of the human voice are contained in the narrow band. Voice sampling using the wideband offers a clearer and fuller-sounding voice representation but at the cost of higher bandwidth requirements.

*B. Echo*

Echo is a result of electrical impedance mismatches in the transmission path. Echo is always present, even in traditional telephony networks, but at a level that cannot be detected by the human ear. The two components that affect echo are amplitude (loudness of the echo) and delay (the time between the spoken voice and the echoed sound). To reduce the annoying effects of the echo phenomenon, a specific system, referred to as echo canceller or suppressor has been considered for a more efficient deployment of the VoIP technology.

*C. Side tone*

Side tone refers to the fact that the telephone allows the speakers to hear their spoken audio in the telephone earpiece. Without side tone, the speaker is left with the impression that the telephone instrument is not working.

*D. Background noise*

It corresponds to the low-volume audio that is heard from the far-end connection to prevent the illusion that the call has been disconnected.

V. CONCLUSION

In this paper, we have presented a deep analysis of the QoS Issues encountering the deployment of the VoIP technology. Firstly, we have presented a brief overview about the basics of the VoIP technology. Then, we have discussed the QoS issues related to the use of the IP networking technology for voice traffic transmission. After that, we have presented the QoS concerns related voice clarity. Finally, we have presented the QoS mechanisms proposed to support voice traffic QoS concerns in terms of voice clarity, voice packet delay, packet delay variation, and packet loss.